\documentclass[fullpage]{article}
\usepackage{multirow,multicol}
\usepackage{color,tabularx,colortbl}
\usepackage{graphicx}
\usepackage{url}
\usepackage{xcolor}
\usepackage[numbers]{natbib}
\usepackage{pdflscape}
\usepackage{afterpage}
\usepackage{rotating}
\usepackage{fullpage}

\begin{document}
\title{Shallow pooling for sparse labels}

\author{
  Negar Arabzadeh
  \and Alexandra Vtyurina
  \and Xinyi Yan
  \and Charles L. A. Clarke
}
\date{}

\date{School of Computer Science, University of Waterloo, Canada}

\maketitle

\begin{abstract}
Recent years have seen enormous gains in core information retrieval tasks, including document and passage ranking. Datasets and leaderboards, and in particular the MS MARCO datasets, illustrate the dramatic improvements achieved by modern neural rankers. When compared with traditional information retrieval test collections, such as those developed by TREC, the MS MARCO datasets employ substantially more queries~---~thousands vs. dozens~--~with substantially fewer known relevant items per query~---~often just one. For example, 94\% of the nearly seven thousand queries in the MS MARCO passage ranking development set have only a single known relevant passage, and no query has more than four. Given the sparsity of these relevance labels, the MS MARCO leaderboards track improvements with mean reciprocal rank (MRR). In essence, the known relevant item is treated as the ``right answer'' or ``best answer'', with rankers scored on their ability to place this item as high in the ranking as possible. In working with these sparse labels, we have observed that the top items returned by a ranker often appear superior to judged relevant items. Others have reported the same observation.

To test this observation, we employed crowdsourced workers to make preference judgments between the top item returned by a modern neural ranking stack and a judged relevant item for the nearly seven thousand queries in the passage ranking development set. The results support our observation. If we imagine a hypothetical perfect ranker under MRR, with a score of 1 on all queries, our preference judgments indicate that a searcher would prefer the top result from a modern neural ranking stack more frequently than the top result from the hypothetical perfect ranker, making our neural ranker ``better than perfect''.

To understand the implications for the leaderboard, we pooled the top document from available runs near the top of the passage ranking leaderboard for over 500 queries. We employed crowdsourced workers to make preference judgments over these pools and re-evaluated the runs. Our results support our concerns that current MS MARCO datasets may no longer be able to recognize genuine improvements in rankers. In future, if rankers are measured against a single answer, this answer should be the best answer or most preferred answer, and maintained with ongoing judgments.
Since only the best known answer is required, this ongoing maintenance might be performed with shallow pooling. When a previously unjudged document is surfaced as the top item in a ranking, it can directly compared with the previous best known answer.
\end{abstract}

\newpage
\section{Introduction}

The last three years have seen a dramatic shift in the state-of-the-art for many core IR tasks. As recently as 2018 it was unclear that neural rankers could outperform non-neural rankers on traditional adhoc document and passage ranking tasks when only content-based, textual features are available~\cite{lin19}. For well over a decade, non-neural learning-to-rank methods have become firmly established in contexts where many non-content features are available, such as web search~\cite{bsr05,jr07}. Unfortunately, attempts to extend these methods to content-based ranking have had a mixed record unless substantial volumes of data training are available~\cite{sb09,yahoo}.

Before 2018,
a state-of-the-art ranker for these core document and passage ranking tasks might employ BM25 followed by a pseudo-relevance feedback method such as RM3, as typified by the open-source Anserini system\footnote{\url{https://github.com/castorini/anserini}} from the University of Waterloo~\cite{yang2018anserini}.
By 2020, a state-of-the-art ranker might employ a dense retriever~\cite{xiong2020approximate,ding2020rocketqa,hofstatter2020improving,khattab2020colbert} followed by one or more transformer-based re-rankers~\cite{nogueira2019passage,nogueira2019multi,han2020learning}.
In 2021, when we we conducted the experiments reported in this paper,
the state-of-the-art could reasonably be represented by RocketQA, a dense retriever which utilizes a dual encoder as well as a cross encoder architecture in order to learn dense representations of queries and passages~\cite{ding2020rocketqa}.
Since then, the state-of-the-art has continued to evolve~\cite{sota0,sota1,sota2}.

Over this four-year period, the IR community has tracked this progress on a number of leaderboards, most notably the MS MARCO\footnote{\url{https://microsoft.github.io/msmarco/}} leaderboards~\cite{marco21short,marco21full}. The MS MARCO project creates test collections focused on deep learning for search. Each test collection is based on a corpus of passages or documents, and comprises a set of training queries, a set of development (i.e., validation) queries, and a set of evaluation (i.e., test) queries. With a few exceptions, each query has one known relevant item in the associated corpus. We call these labeled relevant items ``qrels'' for simplicity\footnote{In established jargon, the word ``qrel'' describes any judgment, which could be relevant or non-relevant, graded or binary, multi-faceted, etc. Here, we use the term exclusively to mean an item judged {\em relevant} on a binary scale. MS MARCO collections do not contain explicitly non-relevant items, so that MS MACRO qrels always indicate a relevant item}. For the training and development sets these qrels are public. For the evaluation sets, the qrels are private. To establish a place on a leaderboard, research groups train and validate rankers with the training and development sets, run the evaluation queries to produce a ranked list for each query, and submit this run to the MS MARCO team.

Since there is often only one qrel per query, these leaderboards use mean reciprocal rank (MRR) as their tracking measure. In effect, a qrel represents the ``right  answer'', and rankers are evaluated on their ability to place this answer as close to the top as possible. This approach stands in contrast to the approach used for many traditional information retrieval evaluations, such as those conducted as part of the long running TREC evaluation exercise\footnote{\url{https://trec.nist.gov}}. In a typical TREC experiment, top-ranked items from each submission are pooled for relevance judgments, so that measures such as NDCG may be applied. Unfortunately, even relatively shallow pooling (e.g., the top three items from each submission to compute NDCG@3) often requires dozens of judgments per query, limiting the number of queries for each experiment. As we will discuss in Section~\ref{sec:marco} the method used to identify MS MARCO qrels supports our view that these qrels do not reflect relevance in the traditional sense, but instead only represent an attempt to identify an answer, and not necessarily the best answer.

In this paper, we focus our attention on the MS MARCO passage retrieval leaderboard. We have no evidence that the concerns raised in this paper would apply to the document retrieval leaderboard, but we also have no evidence that they wouldn't, and we leave that investigation to future work. For the passage retrieval leaderboard the corpus comprises 8.8 million passages extracted from web pages, with queries sampled from the Bing search engine. For this leaderboard, the training set comprises over 500K queries, the development set comprises 6,980 queries, and the evaluation set comprises 6,837 queries. In the development set 6,590 queries (94\%) have only a single qrel and no query has more than 4 qrels. The development set is public while the evaluation set is kept private.

While the passage retrieval leaderboard tracks improvements over both the development and evaluation sets, MRR@10 on the evaluation set provides the official tracking measure.
The first official baseline established on November 1, 2018 used standard methods dating back to the nineties and achieved an MRR@10 of 0.165 on the development set and an MRR@10 of 0.167 on the evaluation set. At the time our experiments were conducted (January-May 2021) MRR@10 had progressed to 0.426 on the development set and 0.439 on the evaluation set~\cite{ding2020rocketqa}. This result was established by RocketQA on September 18, 2020, and it remained state-of-the-art until July 2021.

In attempting to claim a rung on the leaderboard for ourselves, we observed that the top passages returned by our ranker often appeared as good as, or even superior to, the qrels. We are not the only researchers to make this observation. The creators of RocketQA write, ``...we manually examine the top-ranked passages (retrieved by our retriever) that were not labeled as positives in the original MSMARCO dataset, and we find that 70\% of them are actually positives or highly relevant.'' Based on this observation, they trained a model with a dual encoder architecture that is able to perform cross-batch negative sampling, consequently decreasing the probability of selecting false negative in the training step. Compared to other state-of-the-art dense retrievers, their proposed approach focuses on importance of selecting negative samples in the training step. In part, their success depends on explicitly recognizing that an official ``right answer'' may not be the best answer.

We were disturbed by the ramifications of these observations.
Based on these observations,
the current state-of-the-art could be out-performing the qrels.
Imagine an hypothetical perfect ranker that always puts an official qrel at the
topmost rank, scoring an MRR of 1 on these qrels.
If state-of-the-art rankers are surfacing passages in their top ranks that are
superior to the qrels,
the perfect ranking could be viewed as inferior to the state-of-the-art.
If we placed the top result returned by a state-of-the-art ranker side-by-side with the official qrel, which of the two would people prefer? If they would prefer the top result more often, then the state-of-the-art ranking could be viewed as a ``better than the perfect'' result. Can progress be properly measured if rankers are already better than perfect?

In the remainder of the paper, we explore these observations and their ramifications. In Section~\ref{sec:perfect} we describe an experiment to crowdsource comparisons between the top results from a modern neural ranker and the qrels, confirming our observations. Consistent with the method by which the original qrels were created, we employ preference judgments rather than typical pointwise relevance judgments, since our goal is to determine the better answer. Given the quality of current neural rankers, both answers are likely to be relevant in any traditional sense.

In section~\ref{sec:pool} we pool the top passage from available runs on the development set for 500 queries to either select a new answer for these queries or to confirm the qrel as the best answer. These runs were either generated from code in github repositories and checked against the leaderboard, or provided to us by the MS MARCO team. In Section~\ref{sec:leader} we compare runs using these new qrels to examine possible impacts and implications for current leaderboards.

We conclude with suggestions for the future development and maintenance for
datasets with sparse labels.
Since there may be large numbers of unlabeled but relevant items,
we suggest that sparse labels should explicitly represent the best known items,
with preference judgments employed to determine and maintain these best known items.
When rankers surface unjudged items in their top ranks, the qrels can be maintained by
comparing the unjudged items against the qrels, replacing them when the new items are
preferred.

\section{Background}

\subsection{MS MARCO}
\label{sec:marco}
A recent perspective paper by \citet{marco21full} provides a complete exposition on the background and status of the MS MARCO project. That paper carefully and thoroughly addresses many common concerns regarding the MS MACRO datasets, including questions of internal validity, robust usefulness, and the reliability of statistical tests. In this section, we provide only the background required to fully understand the work reported in the current paper. In particular, \citet{marco21full} address concerns raised by \citet{ferrante2021meaningful} who apply measurement theory to draw attention to important shortcomings of established evaluation measures, such as MRR. Many of these measures are not interval scaled, and therefore many common statistical tests are not permissible, and properly these measures should not even be averaged. These concerns are further addressed in a related paper by the same authors~\cite{marco21short}, which we recommend to readers sharing these specific concerns.

In this paper, we focus solely on the process employed to select the qrel for each topic. \citet{marco21full} address this external validity concern as well, writing, ``...there could be quirks of the MS MARCO sparse labeling that pretrained transformer models can learn, giving good performance on MS MARCO sparse labels in the test set, but the improvements would vanish if we relabeled the data with slightly different judging scheme. In that case, the results would be specific to the setup of our study, lacking external validity. We could only claim a real improvement if we think real users have exactly the same quirks as the MS MARCO labels''.

They argue that this concern is addressed by experiments conducted for the TREC Deep Learning Track~\cite{craswell2020overview,DBLP:journals/corr/abs-2102-07662}, which follows the typical methodology of a TREC evaluation exercise. In a traditional TREC-style evaluation exercise, the top items from each submission are pooled to a fixed depth,  perhaps ten or more, and these items are individually judged by assessors with respect to a defined relevance scale. \citet{marco21full} compare the performance of top run from the MS MARCO leaderboard with best TREC run submitted to the TREC Deep Learning Track on a held-out query set from a private MS MARCO leaderboard. The top run from the leaderboard is not as highly ranked when it is evaluated by the comprehensive TREC labels. They hypothesize that some TREC runs might have used previous TREC 2019 labels. Even though the pretrained transformer models can still perform well on TREC labels, this observation could indicate the lack of external validity. 

Commercial search services make attempts to fully judge items to some specified depth $k$. As changes and improvements to commercial rankers cause them to surface previously unjudged items, these items are judged individually according to established guidelines\footnote{Google's search quality rating guidelines provide an example of commercial-level rating guidelines:  \url{https://static.googleusercontent.com/media/guidelines.raterhub.com/en//searchqualityevaluatorguidelines.pdf}}. With complete judgments for the top $k$ items, measures such as NDCG@$k$ may be computed.

The creators of the MS MARCO collections followed a different path, possibly due to the origins of the MS MARCO passage collection as a reading comprehension dataset~\cite{nguyen2016ms}. Queries for the collection were extracted from the logs of the Bing search engine, with an emphasis on queries that took the form of a question. Top documents were retrieved by Bing and 10 candidate passages were automatically extracted from these documents. These passages were shown to an assessor {\bf as a set} and the assessor identified a passage containing an answer. In some cases, two or more passages were selected. These passages became the qrels.

While nothing indicates that these assessors made comparisons between passages when selecting answers, nothing prevented these comparisons, and the interface presented in Figure~1 of \citet{nguyen2016ms}, which presents passages as a list, does nothing to prevent these comparisons. Moreover, assessors were not required or encouraged to identify all passages containing an answer, nor does it appear that they were encouraged to identify the passage containing the best answer. As the authors indicate, ``there are likely passages in the collection that contain the answer to a question but have not been annotated as [such]''. As a result, the labels for MS MARCO datasets can not be treated as traditional relevance labels, which are assessed independently, nor are they complete.

Since it does not appear that the assessors were encouraged to identify the best answer, an important implication for MRR is left unrecognized. Rankers are rewarded for placing qrels as high as possible in the ranking, but if there are better answers in the collection, it becomes possible for a ranker to outperform the qrels by placing better passages above the qrels. In Section~\ref{sec:perfect}, we experimentally test this possibility.

\subsection{Comparative assessment}

As dicussed in the previous section, MS MARCO qrels implicitly identify certain passages as the ``right answer'' and evaluate experimental runs on their ability to place these answers as high in the ranking as possible. As the state-of-the-art improves and better answers are surfaced, we wish to identify these improved answers and replace the qrels with them. At any point in time, a leaderboard should reflect the ability of rankers to place the best known answer at the top of the ranking. Apart from some evaluation exercises on named item finding and similar tasks (e.g., \citet{msc06}) evaluation by the best known answer is relatively rare. Navigational queries provide a major exception, where returning the desired URL in response to a navigational query remains a core function of Web search engines~\cite{Broder02}.

The ongoing identification of the best known answer naturally lends itself to comparative assessment, rather than the typical pointwise relevance assessment employed by many academic and industry evaluation efforts, including TREC. Comparative assessment has a long history in information retrieval, although it has been rarely adopted by any evaluation exercise. In the context of image search, \citet{shao19} demonstrate that showing groups of images to assessors can produce judgments that better reflect user satisfaction, when compared to individual pointwise judgments. This work was continued by \citet{xie20} who used side-by-side preference judgments to identify fine-grained differences between images. In other recent work, \citet{mmst17} estimated relevance magnitudes by showing assessors sequences of eight documents, so that comparisons could be made between them. 

Other efforts to employ comparative judgments for relevance assessment have met with mixed success. \citet{sz20} crowdsourced side-by-side preference judgments for over 100,000 document pairs for runs submitted to an NTCIR evaluation task, but could identify no clear advantage for preference judgments over traditional pointwise judgments. \citet{ymt18} conducted a similar experiment based on the TREC-8 adhoc test collection, concluding that pairwise preference judgments can be cheap, effective, and as reliable as traditional pointwise judgments. Extensive work by Carterette and various collaborator also indicate potential benefits from pairwise preference judgments~\cite{cbcd08,zc10,cp08,cbc08}, including improved assessment speed and accuracy.

An issue frequently raised in this past work, which may be preventing the widespread adoption of pairwise preference judgments, is the perceived need for a greater number of judgments. Exactly $N$ pointwise judgments are required to fully judge a pool of $N$ items, while a quadratic number of judgments might be required to fully judge the same pool with pairwise preferences. If we assume transitivity of preference judgments,  this number might be reduced to less than $N\log N$ judgments. By transitivity,  we mean that if item $A$ is preferred to item $B$ and, item $B$ is preferred to item $C$, we can assume that item $A$ would be preferred to item $C$. However, since judgments are subject to human disagreement and error, transitivity cannot be assumed. While concerns about the need for a large number of judgments may be valid if the goal is computing NDCG, in this paper our focus is identifying and maintaining a single best known item, which can be based on shallow pools, even just the top document, and can be easily maintained by comparing against the current best known item as new items are surfaced.

Recent work from our research group has focused preference judgments on identifying the top-$k$ results~\cite{clarke2020assessing}.
Starting with deeper pools,
that paper proposed a tournament structure to converge on the top-$k$ results,
while minimizing the overall number of preference judgments required.
In the current paper, we identify a single best answer, the top-$1$ result from
shallow pools, so that a tournament structure is not required.
When a previously unjudged document is surfaced as the top item in a ranking,
it can directly compared with the previous best known answer to maintain an overall
best known answer.

Another issue frequently raised in this past work is the lack of validated evaluation measures for preference judgments. \citet{sz20} extend and explore a family of measures first defined by \citet{cbcd08}. These measure reflect the degree to which a ranking agrees with preference judgments. Unfortunately, \citet{sz20} could demonstrate no clear benefit from these measures when compared with traditional pointwise judgments and NDCG. In this paper, since we are treating a qrel as a best known answer, we can employ MRR as our evaluation measure.
Recent work from our research group has proposed other measures for preference judgments~\cite{cvs20,dag} but for a single best answer, MRR remains appropriate~\cite{marco21full}.

\begin{table}[t]
\centering
\begin{tabular}{|l|r|r|}
\hline
 & Number of queries & \% queries \\ \hline
{\bf Category A}: qrel  at first rank & 1868 & 26.76\% \\ \hline
{\bf Category B}: non-qrel at first rank & 5112 & 73.24\% \\ \hline
Total & 6980 & 100.00\% \\ \hline
\end{tabular}
\caption{Agreement between qrels and top results from our neural ranker. A hypothetical perfect ranker would always place the qrel first.}
\label{tab:AB}
\end{table}


\begin{table}[t]
\centering
\begin{tabular}{l|rr|}
\cline{2-3}
& \multicolumn{1}{c|}{{\bf Category A}} & \multicolumn{1}{c|}{{\bf Category B}}\\ \hline
\multicolumn{1}{|l|}{qrel preferred} & \multicolumn{1}{r|}{1,228} & 2,116 \\ \hline
\multicolumn{1}{|l|}{comparison passage preferred} & \multicolumn{1}{r|}{640} & 2,996 \\ \hline
\multicolumn{1}{|l|}{Total} & \multicolumn{1}{r|}{1,868} & 5,112 \\ \hline
\end{tabular}
\caption{
  Results of passage comparison. 
  For Category A queries, the comparison passage is the second passage returned by
  the neural ranker.
  For Category B queries, the comparison passage is the top passage returned by the
  neural ranker.
}
\label{tab:ABC}
\end{table}

\begin{figure}[t]
{\bf 1) query 423878}: {\em is skin cancer genetic}\\
\begin{quote}
{\bf preferred qrel passage 7590792}:
This was a genetic study looking to identify new genetic origins of melanoma. Melanoma is the most serious type of skin cancer. Some of the risk of developing it may be influenced by the genes that you inherit, or your family history. The researchers estimate that around 10\% of people with melanoma have one first-degree relative previously diagnosed.\\
\\
{\bf non-qrel passage 7590794}:
New genetic clues about skin cancer. Skin cancer: Genetic mutations 'warn of risk',\\
\end{quote}

{\bf 2) query 573954}: {\em what are the steps to a waltz dance}\\
\begin{quote}
{\bf preferred non-qrel passage 7403850}:
The basic step for waltz is a box step. It's named after a pattern it creates on the floor (box or square) and forms the foundation of the dance. A box step can be divided into two parts - a forward half box and a backward half box. Each half box has three steps - a step forward or backward, a step to the side, and a step to close the feet together.\\
\\
{\bf qrel passage 7403851}:
Here is the basic waltz steps diagram for the leader. 1  Step forward with the left foot. 2  Right foot step sideways to the right. 3  Bring your left foot next to your right foot.  Step back with the right 1  foot. Step back sideways with the left foot.  Bring your right foot next to your left foot.\\
\end{quote}

{\bf 3) query 764139}: {\em what is ladder move}\\
\begin{quote}
{\bf preferred non-qrel passage 7912169}:
MOV (Move) Ladder Logic Instruction. The Move instruction is a ladder logic rung output instruction that copies the Source value and places a copy in the Destination tag. The Source remains unchanged. The instruction is enabled when the preceding logic is true and disabled otherwise. The values can be constants, tags or any combination.\\
\\
{\bf qrel passage 7912164}:
Up is not the only way forward. Climbing the ladder is the traditional model for career growth, taking a single pathway upward through the corporate hierarchy. However, it's not the only way to move forward.\\
\end{quote}
\caption{
Three arbitrarily selected examples of preference judgments for the experiment reported in Section~\ref{sec:perfect}.
These examples were selected from our logs without prior knowledge of the questions,
passages or outcome.
In each case, we list the preferred passage first.
In the top example, the preferred passage appears more complete than the other passage.
In the middle example, the preferred passage stands on its own, while the alternative refers to an unseen diagram and also only provides steps for the dance leader.
In the bottom example, the passages relate to different interpretations of the query.
All six passages could reasonably be judged relevant under traditional pointwise
assessment.
}
\label{fig:eg}
\end{figure}

\begin{figure}[t]
  \centering
  \label{fig1}
  \includegraphics[width=0.667\linewidth]{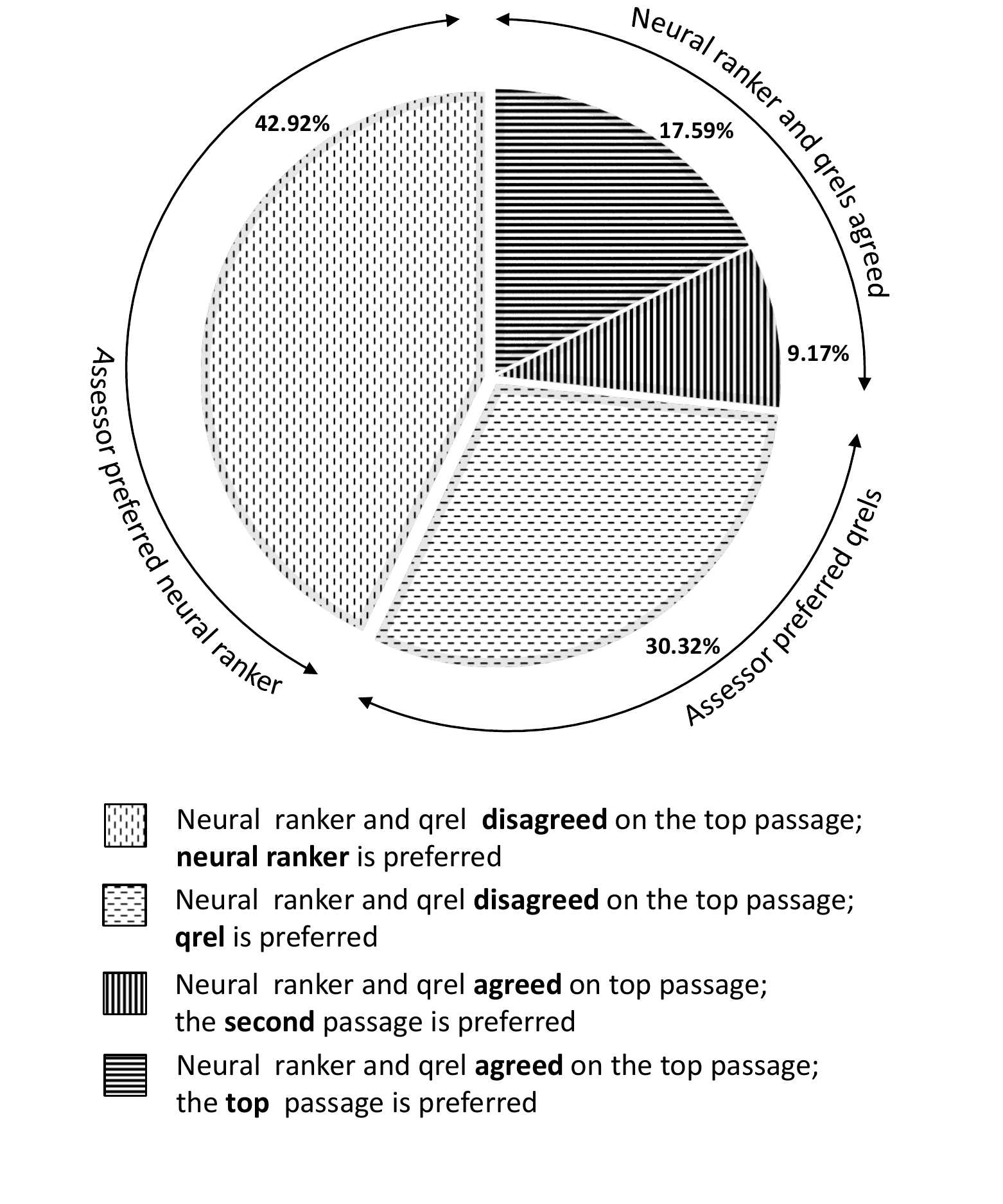}
  \caption{
Results of passage comparison. When the neural ranker and the qrels agree on the top passage, crowdsourced assessors prefer it over the second passage from the neural ranker nearly 65.7\% of the time. When the neural ranker and the qrels disagree, the crowdsourced assessors prefer the top passage from the neural ranker over the qrel for 58.6\% of the pairs.}
\label{fig:pref}
\end{figure}

\section{Better than perfect}
\label{sec:perfect}

For each query, MS MARCO measures the performance of a ranker according to the rank at which it places a single known answer, or qrel, or one of small set of them. The higher the better. MRR@10 is then used to average over a set of queries. As we observe in the previous section, the qrels do not include all answers, nor neccessarily even the best answer. In this section, we further explore the ramifications of this observation.

We compare the top passages returned by a representative neural ranker with the qrels used to evaluate experimental runs. If the qrels were the best answers, we would expect a preference for qrels over other passages. Since both the qrels and our subsequent preference judgments depend on error-prone human assessments, this preference won't be universal, but we would certainly expect qrels to be preferred the majority of the time.

\afterpage{
\clearpage
\begin{landscape}
\begin{table}[p]
\begin{center}
\begin{tabular}{|l|r|r|r|r|r|r|}
\hline
\multicolumn{1}{|c|}{} & \multicolumn{1}{c|}{} & \multicolumn{4}{c|}{MRR@10} &  \\ \cline{3-6}
\multicolumn{1}{|c|}{\multirow{-2}{*}{\begin{tabular}[c]{@{}c@{}}Run\\ Name\end{tabular}}} & \multicolumn{1}{c|}{\multirow{-2}{*}{\begin{tabular}[c]{@{}c@{}}Leaderboard\\ Rank\end{tabular}}} & \multicolumn{1}{c|}{\cellcolor[HTML]{FFFFFF}\begin{tabular}[c]{@{}c@{}}Evaluation Set\\ All queries\end{tabular}} & \multicolumn{1}{c|}{\cellcolor[HTML]{FFFFFF}\begin{tabular}[c]{@{}c@{}}Development Set \\ Original qrels\\ All Queries\end{tabular}} & \multicolumn{1}{c|}{\cellcolor[HTML]{FFFFFF}\begin{tabular}[c]{@{}c@{}}Development Set \\ Original qrels\\ Selected Queries\end{tabular}} & \multicolumn{1}{c|}{\cellcolor[HTML]{FFFFFF}\begin{tabular}[c]{@{}c@{}}Development Set \\ Preference qrels\\ Selected Queries\end{tabular}} & \multirow{-2}{*}{Reference} \\ \hline
\rowcolor[HTML]{FFFFFF} 
Perfect ranking &  & 1 & 1 & 1 & 0.3320 & \\ \hline
\rowcolor[HTML]{FFFFFF} 
A & 3 & 0.4190 & 0.4205 & 0.4013 & 0.5069 & {\em not available} \\ \hline
\rowcolor[HTML]{FFFFFF} 
B & 4 & 0.4080 & 0.4213 & 0.3953 & 0.5638 &  \citet{pradeep2021expando} \\ \hline
\rowcolor[HTML]{FFFFFF} 
C & 5 & 0.4070 & 0.4210 & 0.3881 & 0.5827 &  \citet{han2020learning}\\ \hline
\rowcolor[HTML]{FFFFFF} 
D & 9 & 0.4010 & 0.4118 & 0.3863 & 0.5036 &  {\em not available}\\ \hline
\rowcolor[HTML]{FFFFFF} 
E & 10 & 0.4000 & 0.4074 & 0.3722 & 0.4955 & {\em not available}\\ \hline
\rowcolor[HTML]{FFFFFF} 
F & 11 & 0.3990 & 0.4079 & 0.3697 & 0.5703 & \citet{hofstatter2020improving} \\ \hline
\rowcolor[HTML]{FFFFFF} 
G & 12 & 0.3950 & 0.4046 & 0.3658 & 0.5199 & \citet{han2020learning} \\ \hline
\rowcolor[HTML]{FFFFFF} 
H & 13 & 0.3940 & 0.3998 & 0.3606 & 0.5617 & {\em not available} \\ \hline
\rowcolor[HTML]{FFFFFF} 
I & 30 & 0.3790 & 0.3904 & 0.3541 & 0.5464 & \citet{nogueira2019multi} \\ \hline
\rowcolor[HTML]{FFFFFF} 
J & 34 & 0.3760 & 0.3880 & 0.3442 & 0.4985 & \citet{liu2021openmatch} \\ \hline
\rowcolor[HTML]{FFFFFF} 
K & 81 & 0.3090 & 0.3181 & 0.2984 & 0.3663 & \citet{hofstatter2019effect} \\ \hline
\rowcolor[HTML]{FFFFFF} 
L & 86 & 0.2940 & 0.3039 & 0.2779 & 0.4666 & \citet{zhan2020repbert} \\ \hline
\rowcolor[HTML]{FFFFFF} 
M & 92 & 0.2770 & 0.2901 & 0.2681 & 0.3251 & \citet{hofstatter2019effect} \\ \hline
\rowcolor[HTML]{FFFFFF} 
N & 93 & 0.2720 & 0.2768 & 0.2550 & 0.3257 & \citet{nogueira2019doc2query}\ \\ \hline
\rowcolor[HTML]{FFFFFF} 
O & 110 & 0.2180 & 0.2216 & 0.1955 & 0.2403 &  \citet{nogueira2019document} \\ \hline
\rowcolor[HTML]{FFFFFF} 
P & 115 & 0.1860 & 0.1873 & 0.1536 & 0.2033 &  \citet{yang2017anserini}\\ \hline
\end{tabular}
\end{center}
\caption{MS MARCO passage ranking development runs used in our experiments, ordered according to the MS MARCO leaderboard at the time of writing. These were gathered from a variety of sources, as described in Section~\ref{sec:pool}. The table lists MRR@10 values on several sets of qrels. The ``Selected Queries'' are described in Section~\ref{sec:pool}. The ``Preference qrels'' are described in Section~\ref{sec:leader}.}
\label{tab:main}
\end{table}
\end{landscape}
\clearpage
}

\subsection{Method}

As a representative neural ranker, we employ the ranking stack described by \citet{nogueira2019multi} as provided by their github repository. This ranking stack utilizes Anserini as a sparse first-stage retriever, which is followed by two BERT-based re-rankers, which they call MonoBERT and DuoBERT. The former was introduced by \citet{nogueira2019passage} as a pointwise, second-stage re-ranker. The re-ranked list produced by Monobert feeds the third-stage DuoBERT pairwise re-ranker to generate the final ranking. Although this ranking stack was well over a year old and had sunk to 30th place on the leaderboard at the time our experiments were conducted in early 2021, it remains representative of the technology underlying the dramatic improvements of the past few years. We choose this ranking stack for our experiments for no other reason that we happened to be working with it when we observed the apparent superiority of the top passages it returned.

For our experiments, we split the queries in the development set as follows:
\begin{enumerate}
    \item {\bf Category A}: Queries for which the top passage and the qrel are the same.
    \item {\bf Category B}: Queries for which the top passage and the qrel are different.
\end{enumerate}
While 94\% of the queries in the development set have a single qrel, some have up to four. For these queries, we arbitrarily selected the first qrel in the file as the qrel used for this comparison. Alternatively, for these queries we could have selected the highest ranked qrel returned by the ranker or conducted multiple comparisons. Our approach keeps the selection of the qrel independent of the ranker and avoids multiple comparisons on the same query.

Table~\ref{tab:AB} shows the split between the categories for the 6,980 queries in the development set. For the Category A queries, where the ranker and qrels agree, we would expect this passage to be generally preferred over other passages. For comparison purposes, we choose the second passage returned by our ranker. Even though this passage is highly ranked, we would still expect the top passage to be preferred when the two passages are placed side-by-side. For the Category B queries, we compare the top passage with the qrel. Altogether we have a pair for each query, which can be compared side-by-side by a human assessor to determine which represents the better result.

For assessment, we crowdsourced comparisons on the Amazon Mechanical Turk platform. To build tasks for crowdsourced workers, we grouped comparisons into sets of ten, which were combined with three manually constructed test comparisons, which helped to assure quality. These test comparisons were always between a known relevant passage (i.e., a qrel) and an obviously off-topic and non-relevant passage. Data from workers failing one or more of these test comparisons was excluded from our experiment and the task was repeated by a different worker. For each task we randomized the ordering of the pairs, as well as the right-to-left ordering of passages in the pairs. A full task comprises these thirteen pairs, which are presented one at a time to a worker along with the associated query. For each pair, the worker was asked to indicate the passage that ``best answers the question.''

The Mechanical Turk platform allows tasks to be restricted to workers that satisfy specified criteria. Since the MS MARCO passages and queries are taken from an EN-US environment, we restricted tasks to workers who were located in the US, with more than ten thousand approved tasks and an approval rate of at least 97\%. In addition, we excluded workers who had previously failed the test comparisons for one our tasks. For each task containing 13 pairs, we paid \$2.00, plus a fee of \$0.40 to Amazon. Total cost was \$1,720, include pilot tests, debugging, and other minor costs.

This data labeling process was reviewed and approved by our institutional review board. After accepting a task, workers were shown a consent form and were required to provide consent before undertaking the task. Workers were given an option to exit the task at any point. Workers exiting before completing the task were paid a prorated amount. As required by our review board, the rate of pay for a task was estimated to provide compensation consistent with our local minimum wage.
As part of the ethics review process we provided an example task. This example task was completed by the reviewer, who specified a minimum amount we could pay for the task.
Assessed pairs (without identifying information) were approved for release after completion of the experiment.
Examples of questions, qrels, passages, and preferences are provided in Figure~\ref{fig:eg}.

\begin{figure}[p]
  \centering
  \includegraphics[width=6.0in,keepaspectratio]{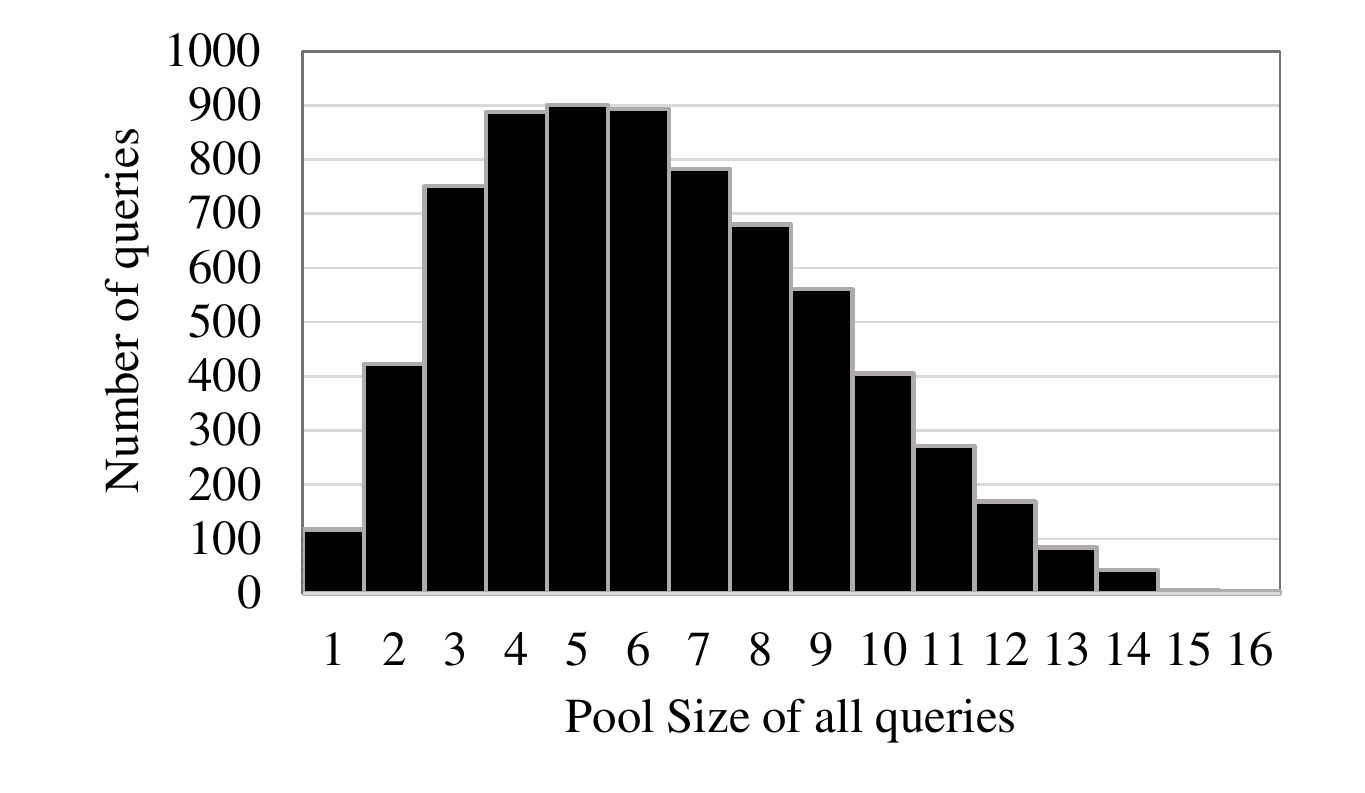}
  \caption{Sizes of preference judgment pools for all 6,980 queries in the MS MARCO passage retrieval development set pooled over all 16 available runs.}
  \label{fig:bigpools}
  \vspace{-2em}
\end{figure}

\begin{figure}[p]
  \centering
  \vspace*{1cm}
  \includegraphics[width=6.0in,keepaspectratio]{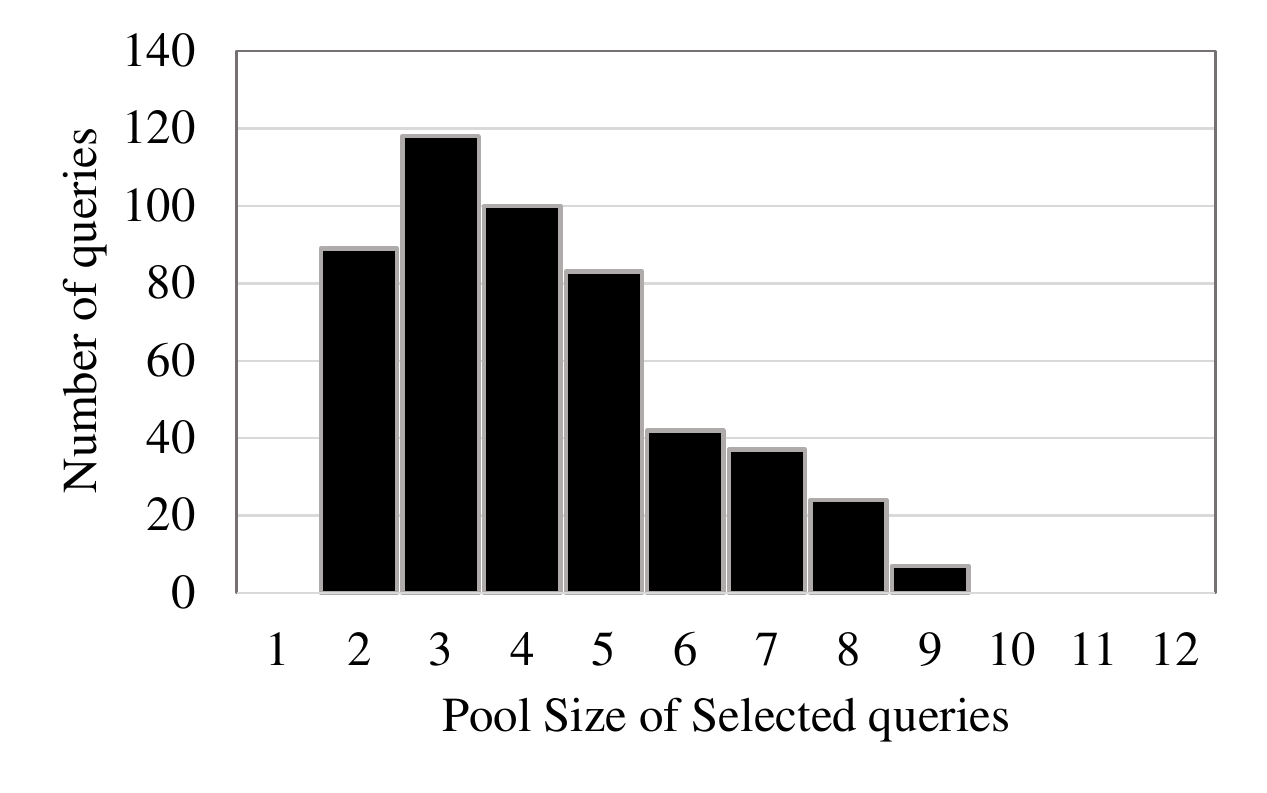}
  \caption{Sizes of preference judgment pools for 500 selected queries pooled over 12 selected runs.}
  \label{fig:smallpools}
\end{figure}

\begin{figure}[t]
    \centering
    \includegraphics[width=7.0in,keepaspectratio]{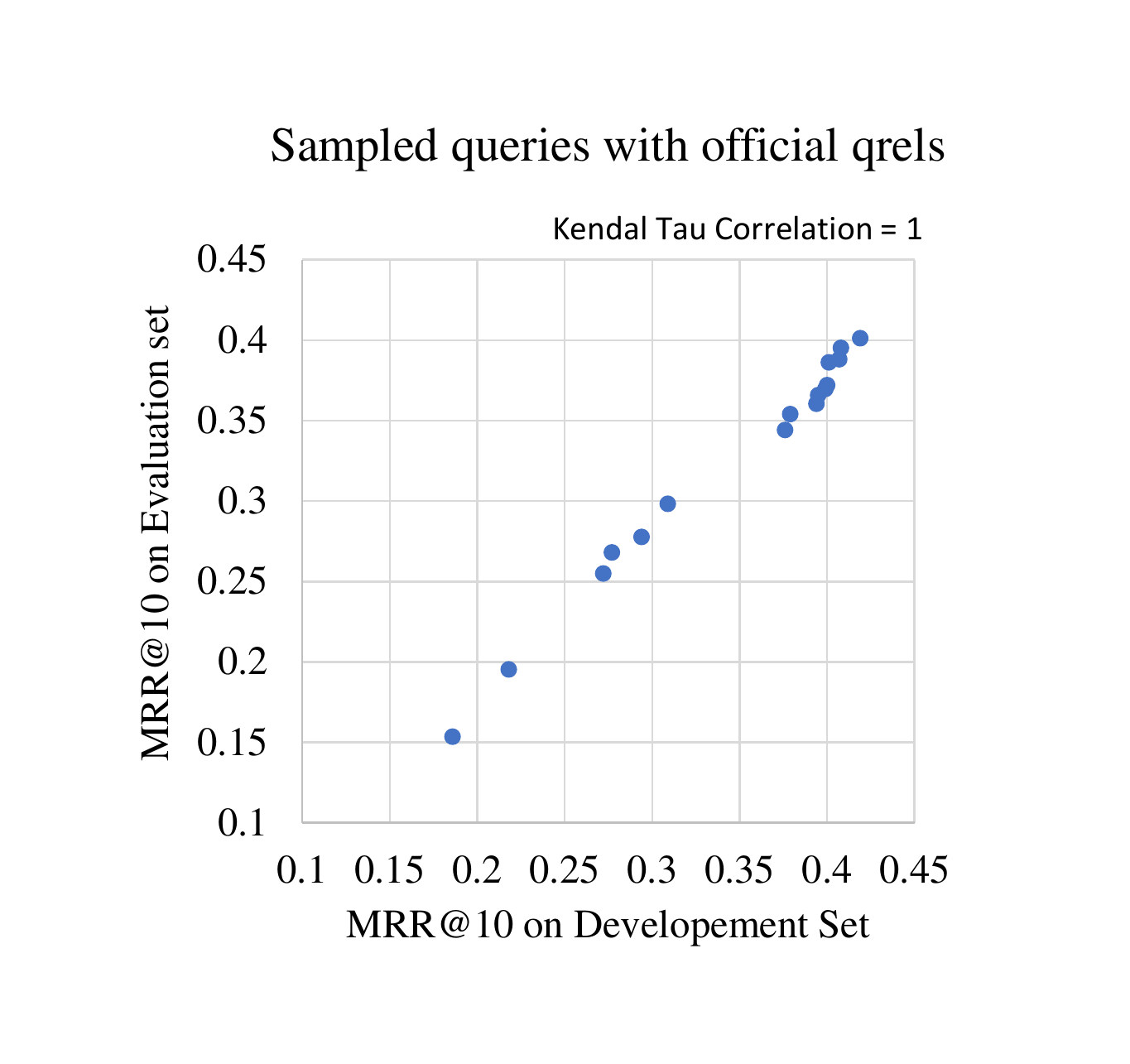}
    \caption{MRR@10 on Evaluation set vs Development set with Official qrels on 500 sampled queries. The rankings are identical suggesting that results on the 500-query sample can be generalized to the development set leaderboard.}
    \label{fig:evalvsdev}
\end{figure}

\subsection{Results}

Figure~\ref{fig:pref} and Table~\ref{tab:ABC} show the results of the crowdsourced passage comparison. For Category A queries, where the neural ranker returns the qrel as the top passage, the assessors agreed that this passage was preferable to the passage ranked second by the neural ranker roughly two-thirds of the time (18\% vs. 9\%). Given that this second-place passage is itself not unlikely to provide a reasonable response to the query, this outcome provides some assurance regarding both the crowdsourcing process and the quality of the ranker. If we assume that the qrel/top passage is in fact the best passage in the collection overall, this result also provides a rough estimate of crowdsourcing error at 33\%.

For the Category B queries, the top passage from the neural ranker was preferred over the qrel for nearly 59\% of the queries (43\% vs. 30\%). Since the result on the Category A queries provides us with some confidence in the crowdsourcing process, this result suggests that the neural ranker is out-performing a hypothetical perfect ranker that always returned the qrel in the top rank. It is in this sense we claim that current neural rankers may already be ``better than perfect'' on the MS MARCO leaderboards. In the following sections we consider the potential impact of this result on our understanding of the the state-of-the-art.

\afterpage{
\clearpage
\begin{landscape}
\begin{table*}[]
\begin{center}
\begin{tabular}{|l|r|r|r|r|r|r|r|r|r|r|r|r|}
\hline
 & \multicolumn{1}{l|}{Perfect Ranking} & \multicolumn{1}{l|}{A} & \multicolumn{1}{l|}{B} & \multicolumn{1}{l|}{C} & \multicolumn{1}{l|}{D} & \multicolumn{1}{l|}{E} & \multicolumn{1}{l|}{F} & \multicolumn{1}{l|}{G} & \multicolumn{1}{l|}{H} & \multicolumn{1}{l|}{I} & \multicolumn{1}{l|}{J} & \multicolumn{1}{l|}{L} \\ \hline
Perfect Ranking & \multicolumn{1}{l|}{} & 49.3\% & \textbf{60.8\%} & 58.6\% & 51.4\% & 49.7\% & \textbf{59.0\%} & 55.8\% & \textbf{58.7\%} & 55.6\% & 51.8\% & 51.0\% \\ \hline
A & 50.7\% & \multicolumn{1}{l|}{} & 55.2\% & 58.0\% & 48.4\% & 45.7\% & 53.1\% & 53.8\% & 55.3\% & 52.7\% & 48.6\% & 45.6\% \\ \hline
B & \textbf{39.2\%} & 44.8\% & \multicolumn{1}{l|}{} & 51.5\% & 44.9\% & 44.4\% & 49.0\% & 48.1\% & 52.0\% & 51.2\% & 45.1\% & \textbf{41.1\%} \\ \hline
C & 41.4\% & 42.0\% & 48.5\% & \multicolumn{1}{l|}{} & 42.4\% & 42.5\% & 49.2\% & 34.5\% & 51.9\% & 47.9\% & 39.8\% & \textbf{40.5\%} \\ \hline
D & 48.6\% & 51.6\% & 55.1\% & 57.6\% & \multicolumn{1}{l|}{} & 46.4\% & 54.6\% & 53.1\% & 55.8\% & 52.1\% & 50.6\% & 46.7\% \\ \hline
E & 50.3\% & 54.3\% & 55.6\% & 57.5\% & 53.6\% & \multicolumn{1}{l|}{} & 54.2\% & 51.5\% & 54.8\% & 51.1\% & 47.6\% & 48.1\% \\ \hline
F & \textbf{41.0\%} & 46.9\% & 51.0\% & 50.8\% & 45.4\% & 45.8\% & \multicolumn{1}{l|}{} & 44.3\% & 55.6\% & 49.8\% & 42.5\% & 42.0\% \\ \hline
G & 44.2\% & 46.2\% & 51.9\% & 65.5\% & 46.9\% & 48.5\% & 55.7\% & \multicolumn{1}{l|}{} & 56.3\% & 55.8\% & 44.1\% & 43.3\% \\ \hline
H & \textbf{41.3\%} & 44.7\% & 48.0\% & 48.1\% & 44.2\% & 45.2\% & 44.4\% & 43.7\% & \multicolumn{1}{l|}{} & 47.5\% & 43.8\% & 42.0\% \\ \hline
I & 44.4\% & 47.3\% & 48.8\% & 52.1\% & 47.9\% & 48.9\% & 50.2\% & 44.2\% & 52.5\% & \multicolumn{1}{l|}{} & 44.1\% & 42.2\% \\ \hline
J & 48.2\% & 51.4\% & 54.9\% & 60.2\% & 49.4\% & 52.4\% & 57.5\% & 55.9\% & 56.3\% & 55.9\% & \multicolumn{1}{l|}{} & 49.7\% \\ \hline
L & 49.0\% & 54.4\% & \textbf{58.9\%} & \textbf{59.5\%} & 53.3\% & 51.9\% & 58.0\% & 56.7\% & 58.0\% & 57.8\% & 50.3\% & \multicolumn{1}{l|}{} \\ \hline
\hline
Wins
& \multicolumn{1}{c|}{2} 
& \multicolumn{1}{c|}{4}
& \multicolumn{1}{c|}{8}
& \multicolumn{1}{c|}{10}
& \multicolumn{1}{c|}{3}
& \multicolumn{1}{c|}{2}
& \multicolumn{1}{c|}{8}
& \multicolumn{1}{c|}{6}
& \multicolumn{1}{c|}{11}
& \multicolumn{1}{c|}{8}
& \multicolumn{1}{c|}{3}
& \multicolumn{1}{c|}{1}
\\
\hline
\end{tabular}
\end{center}
\caption{Win ratios when runs are directly compared according to their top documents. Percentages indicate the frequency that the run in the column beat the run in the row. Bolded numbers indicate significant differences under a binomial test with $\alpha = 0.05$, after a Bonferroni correction. Under this measure, three runs would outperform a perfect ranking under the official qrels. The last row indicates the number of times the run won. Runs are ordered according to their position on the official leaderboard.}
\label{tab:win}
\end{table*}
\end{landscape}
\clearpage
}

\section{Shallow pooling}
\label{sec:pool}

In the previous section, we compared the top documents returned by a single neural ranker to the qrels. When they disagreed, the top passage from the neural ranker was preferred more often. In this section we extend this comparison to encompass a set of top submissions from the leaderboard. For each query, we create a pool of candidate best answers by taking the top passage from each submission, plus the qrel. The passages in each pool are paired and judged by crowdsourced workers. We then directly compare runs in terms of their top documents.

\subsection{Method}

We base our experiments on the runs listed in Table~\ref{tab:main}. Like the previous experiment, this experiment uses the development set because qrels for evaluation queries are not publicly released.  The leaderboard rank indicates position of a run as is was in early 2021, when the research effort reported in this paper was first undertaken. The table lists official MRR@10 scores on both the development queries and evaluation queries, and the table is ordered by MRR@10 on evaluation set. The table also includes a preference MRR@10 using the qrels created by the process described in Section~\ref{sec:leader}.

Since runs are not provided through the MS MARCO leaderboard, we created or obtained these runs from a variety of sources. In some cases, as referenced, the runs could be re-generated from code in github repositories and checked against the leaderboard. In other cases, the runs were provided by to us by the MS MACRO team for the purposes of this paper. We sincerely thank the MS MACRO team for responding positively to our ``cold call'' and providing those runs on the development set that were easily available to them. We focused our efforts to recover runs on the top of the leaderboard, and were able to recover four of the top ten and eight of the top 20. Unfortunately, we were not able to recover the top RocketQA run.

We pooled the top passage from each run, plus the qrel if it did not appear among these top passages. For this experiment, we continue to use a single qrel per query, as described in the previous section. Figure~\ref{fig:bigpools} shows the size of these pools. For 117 queries the pool size is one, i.e., all rankers returned the qrel as the top document. The average pool size is 6.23 passages, with a median pool size of 6 passages. If we paired the documents in these pools and judged each pair once, it would require 141,887 comparisons and cost over \$34,000.

Since this experiment is exploratory, rather than a complete evaluation of the runs themselves, we reduced our costs in two ways. First, we judged a random subset of 500 queries, rather than the full set of queries, as explained below. Second, we restricted judging to the available runs at the top of the leaderboard, i.e, runs A to K, which are more likely to surface documents that outperform the qrels. These runs all have an evaluation MRR@10 between 0.367 and 0.419, while the next-highest run has an MRR@10 of 0.309. In addition, we included run L, which exhibited an interesting property on the pairs from the previous experiment. For this run, when both passages in a judged pair appeared in its top-10 ranking, they were ordered consistently with the preferences more often than any other run, with 63.07\% agreement.

We build new pools for this restricted set of runs, and then selected 500 queries at random for which the pool size was at least two. Figure~\ref{fig:smallpools} shows the size of these pools. The average pool size is 6.32 passages, with a median pool size of 4 passages. Sampling down to these 500 queries from the development set has no impact on the leaderboard ranking, as shown in Figure~\ref{fig:evalvsdev}. This figure plots the MRR@10 on the evaluation set, as shown on the leaderboard, against the MRR@10 on the 500 sampled queries from the development set. The values are shown in Table~\ref{tab:main} in columns~3 and~5. The rankings are identical.

Judging each pair once requires 4,210 comparisons. We crowdsourced these comparisons on Mechanical Turk following the procedure described in the previous section, for an actual cost of \$1,022, including pilot tests, debugging, and other minor costs. Assessed pairs (without identifying information) were approved by our institutional review board for release after completion of the experiment. However, since these pairs include passages surfaced by runs provided to us by the MS MARCO team strictly for this experiment, their general release is not possible.

\subsection{Results}

Just as we did in Section~\ref{sec:perfect}, we directly compare pairs of runs with the preference judgments. For each pair of runs, we compute a {\em win ratio} for the top passages from each run. When the top documents differ, the win ratios indicates how often one run is preferred over another.

The results are shown in Table~\ref{tab:win}. Win ratios compare the columns to the rows, so that a value above 50\% indicates that the run in column beats the run in row more often that not. Bolded numbers indicate significant differences under a binomial test with $\alpha = 0.05$, after a Bonferroni correction. Even under the conservative Bonferroni correction three runs significantly outperform a perfect ranking under the official qrels. The last row indicates the number of wins for the run in that column. Run H is particularly interesting since it wins against all other runs. Unfortunately, we know little about this run, since there is no associated reference.

\section{Leaderboard impact}
\label{sec:leader}

MS MARCO queries can have as little as one qrel, i.e., only one known relevant item. We hypothesize that the MS MARCO evaluation methodology, and its associated leaderboards, depend on this item being the best item, rather than just any relevant item. MRR rewards rankers for placing this item as high as possible in its ranking. Employing MRR as the primary evaluation measure makes the implicit assumption that placing this item above all other items is always the correct thing to do. The experiments in previous sections contradict this assumption, demonstrating that some runs on the leaderboard can be considered ``better than perfect'' according to the existing qrels. In this section, we explore the possibility that these experimental results raise practical concerns, with the potential for impacting leaderboards and our notion of the state-of-the-art.

In this section, we derive new preference-based qrels from the preference judgments described in the previous section. Using these preference qrels, we compute new MRR@10 values for the 16 available runs. We compare the results with MRR@10 values computed using the original qrels.

\subsection{Method}

We convert preference judgments to qrels by treating the preference judgments as a tournament. If a single passage wins the most pairings, we designate that passage to be the ``preference qrel'' for that query. For queries where multiple passages were tied for first place, we eliminate the losing passages and repeat the process with the first-place passages, until we have a single ``preference qrel''. For 46 of the 500 selected queries, it was not possible to designate a single ``preference qrel'' due to a cycle between three passages. Overall we have 592 qrels in the preference qrel set for the 500 selected queries. The preference qrels win 1598 or 91.8\% of their pairings, while the original qrels win 809 or 46.6\% of their pairings. For the remainder of the experimental results in this section, we use all of the original qrels, not just one for each query. For the 500 selected queries, there are 528 qrels in this original qrel set.

\subsection{Results}

\begin{figure}[t]
    \centering
    \includegraphics[width=4.0in,keepaspectratio]{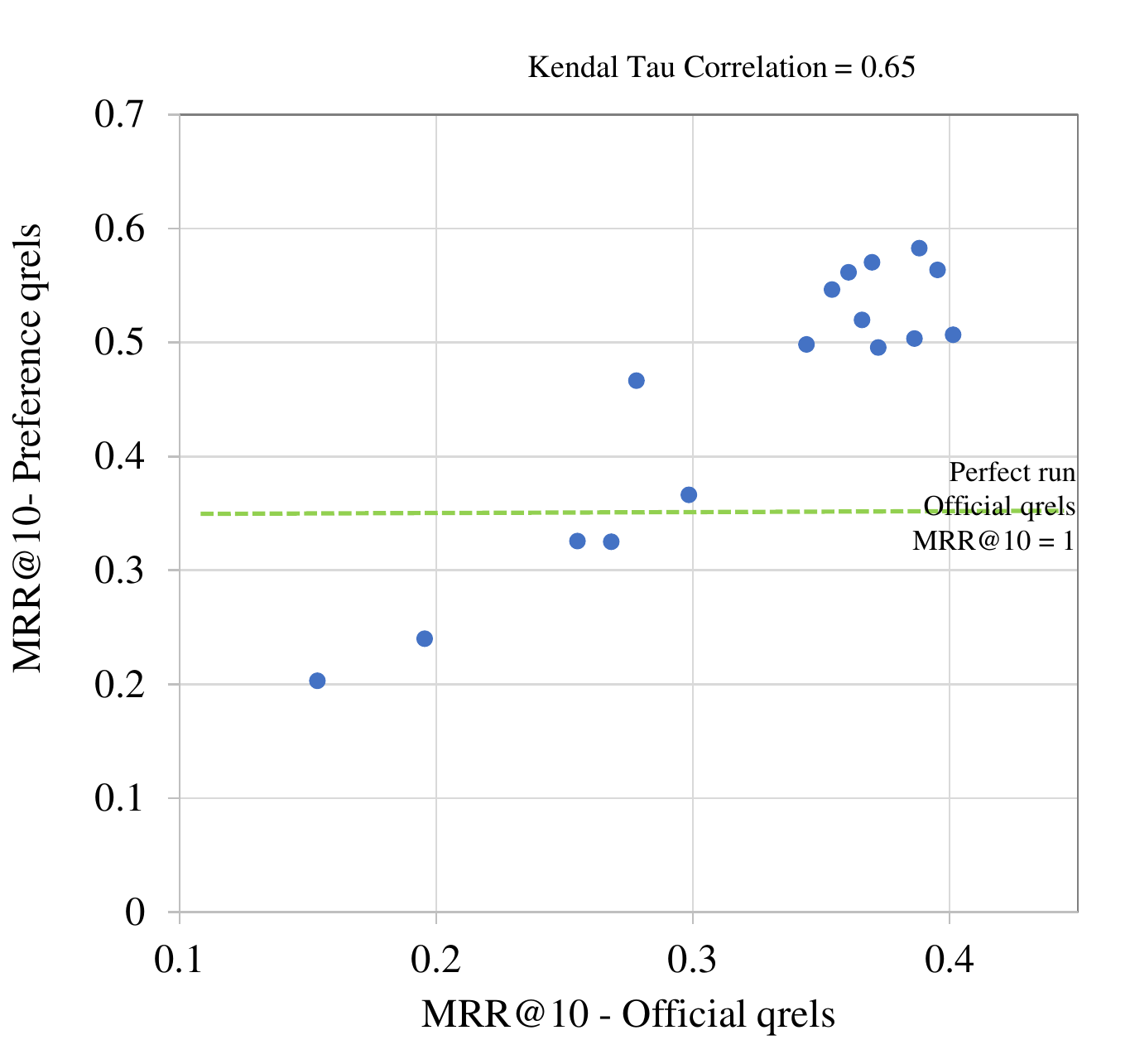}\\
    \includegraphics[height=2.2in,keepaspectratio]{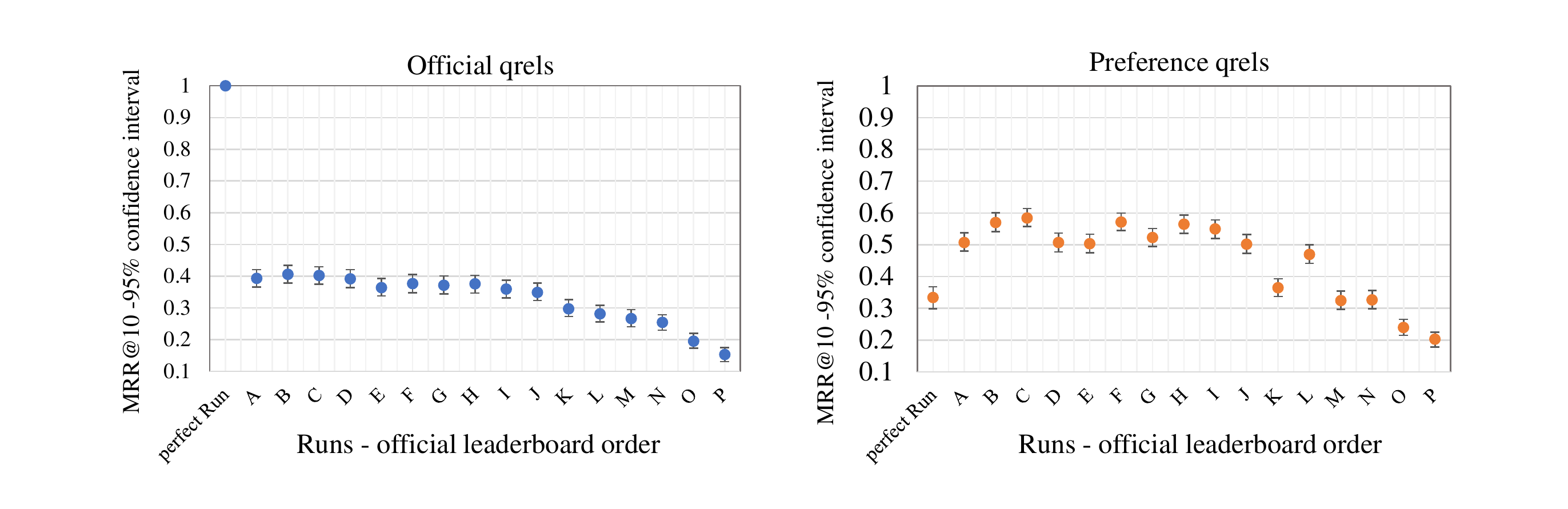}
    \caption{Shallow pooling of top documents from available MS MARCO passage retrieval runs followed by crowdsourced preference judgments to create new qrels for a sample of 500 queries. A perfect ranking under the official qrels, with an MRR@10 of 1, performs poorly under these new preference qrels. Changes in the order of top runs raise concerns about the ability of the official qrels to recognize improvement in the state of the art. The bottom plots show 95\% confidence intervals. Runs are ordered according to the official leaderboard.}
    \label{fig:main}
\end{figure}

The results are given in Figure~\ref{fig:main}.
For these results, we use all 16 runs available, including runs K and M-P, which were not included in the pools for the previous experiment. These runs all placed below the top 80 on the leaderboard at the time our experiments were completed in May 2021.

The first graph in Figure~\ref{fig:main} plots MRR@10 with the preference qrels vs.\ MRR@10 with the original qrels. Although there is a correlation between the two measures over the plot as a whole, with Kendall's $\tau = 0.65$, at the top rungs of the leaderboard, the relative order of the runs changes dramatically. The green dashed line indicates the MRR@10 of the perfect ranking under the official qrels, which scores MRR@10 = 0.3320 on the preference qrels.

The bottom plots order the runs according to the official leaderboard ranks, and show 95\% confidence intervals. On the official qrels (second plot) the runs adhere closely to the leaderboard ranking. On the preference qrels, there are noticeable changes in the ranking. In particular, run A drops below runs B and C, while runs D and E drop below F. 

Run C is the best performing run on the preference qrels, but third on the original qrels. This run is described by \citet{han2020learning} who argue that finetuning a classification model with the aim of deciding whether a document is relevant to a query or not, is not a suitable approach for a ranking task. Instead, they employed a learning-to-rank algorithm on a pair-wise and list-wise basis which learns to differentiate relevance for document pairs or optimize the list as a whole, respectively. Unlike most BERT-based methods, this framework builds a LTR model through fine-tuning representation of query-document pairs and demonstrates the potential of combining ranking losses with BERT representations, especially for passage ranking. Since this run takes a very different approach than most runs at the top of the leaderboard, we hope that a promising direction for future progress has not been missed.

On the preference qrels, run L (\citet{zhan2020repbert}) noticeably improves against runs with a similar MRR@10 on the original qrels. As mentioned previously, this run exhibited an interesting property on the pairs from the previous experiment. When both passages in a judged pair appeared in its top-10 ranking, they were ordered more consistently with the preferences more frequently than any other run. To check that this relative improvement was not an artifact of the run's inclusion in the pool, we re-calculated the preference qrels without the passages it contributed. While the exact numbers, changed, run~L continued to show this relative improvement.

\section{Concluding Discussion}

The MS MARCO leaderboards evaluate submissions with sparse labels indicating relevance (``the qrels'') and Mean Reciprocal Rank (MRR). Imagine a state-of-the-art ranker that routinely surfaces in its top ranked items that a person would prefer to these qrels. Perhaps in comparison to the previous state-of-the-art it also places the qrels higher in the ranking on average, but still below the superior items, so that its measured MRR is larger. To recognize further improvements to the state-of-the-art we are depending on this property~---~that improved rankers will place qrels higher and higher. But in this thought experiment, if improved rankers are surfacing more and more items that are superior to the qrels, these items will tend to push the qrels down, lowering measured MRR. Our results suggests that this phenomenon, or a similar phenomenon, may already be occurring on the passage ranking leaderboard. 

While this paper may appear critical of the MS MARCO effort, the opposite is true. By enabling evaluations with larger query sets and sparser labels, MS MARCO represents a major advance in information retrieval evaluation. With some exceptions, most previous evaluation efforts make an implicit assumption that the judgment set is sufficiently complete. Measures such as NDCG@$k$ require a gain value for each ranked item down to depth $k$. Unjudged items are typically assumed to provide zero gain~\cite{sakai07}. As new retrieval methods surface unjudged items with gain that would equals or exceed that of judged items, significant improvements can be missed~\cite{ycmmc20}. Avoiding the negative impact of unjudged items requires an ongoing commitment to judge all newly surfaced items down to depth $k$.

In contrast, MS MARCO identifies an overall best item, or perhaps several such items, and evaluates rankers by their ability to place these items as high as possible in their rankings. MRR is the primary evaluation measure, not NDCG. As unjudged items are surfaced, we can maintain this set of best items with comparative judgments, instead of traditional pointwise judgments. For example, we might present an assessor with a list of items and ask which is best. The original development of the MS MARCO collection essentially followed this approach, although they did not explicitly request the best answer. Alternatively, we might employ side-by-side preference judgments, as we did in this paper.

Comparative judgments allow for finer distinctions between items than is normally possible with pointwise judgments, and there is a growing body of research literature recognizing and exploring preference judgments for this reason~\cite{sz20,zc10,cbcd08,ymt18}. For example, \citet{xie20} recognize the potential of preference judgments for image search, where absolute relevance grades are difficult to define and images must be assessed on intangible factors such as aesthetic appeal~\cite{shao19}. In addition, preference judgments offer improved assessor agreement and require less time, when compared to pointwise judgments. If the top items returned by modern neural rankers are all highly relevant in the traditional sense, preference judgments allow us to make finer distinctions between them by comparing items side by side.

We provide evidence that the performance of modern neural rankers may already exceed the performance of a hypothetical perfect ranker on the MS MARCO passage ranking leaderboard under the official qrels. Over the 6,980 queries comprising the MS MARCO passage ranking development set, our crowdsourced workers more often prefer the top passage returned by a neural ranker to a judged relevant passage. If we imagine a hypothetical perfect ranker that always returned a judged relevant passage at rank 1, achieving an MRR@10 of 1, the top passage from the neural ranker would be preferred more often, making the neural ranker in this sense ``better than perfect''.
This outcome has implications for measuring further improvements to the state-of-the-art. To provide some sense of these implications, we pooled the top passage from available experimental runs for over 500 queries and employed crowdsourced workers to make pairwise preference judgments between them. Again, we see that the results of neural rankers can be ``better than perfect''. When we construct a new set of qrels from these judgments, we see shifts in the leaderboard which raise concerns about its ability to continue tracking improvements.

The MS MARCO effort has successfully fostered substantial new research on the core informational retrieval tasks of adhoc passage and document ranking, allowing researchers to demonstrate unprecedented improvements in the state-of-the-art~\cite{marco21full,marco21short}. Unfortunately, the lack of an ongoing maintenance plan for the MS MARCO judgment set may hinder further improvements. As a maintenance plan for MS MARCO and similar efforts, we suggest regular pooling and judging of top documents to maintain a set of known best items for queries. In order to identify best items, comparative judgments allow finer distinctions between items to be recognized, which might otherwise might be missed with traditional pointwise judgments.

\bibliographystyle{plainnat}
\bibliography{shallow} 

\end{document}